\documentclass[reprint,aps,prl,twocolumn,amsmath,amssymb,superscriptaddress,floatfix,footinbib,longbibliography]{revtex4-2}

\usepackage{graphicx}   
\usepackage{dcolumn}    
\usepackage{bm}         

\usepackage{epsfig}
\usepackage{epstopdf}
\usepackage[T1]{fontenc}
\usepackage[utf8]{inputenc}
\usepackage{lmodern}
\usepackage[english]{babel}
\usepackage{lipsum}
\usepackage{ae}
\usepackage{siunitx}
\usepackage{amsmath,amssymb}

\usepackage{tikz}
\usetikzlibrary{arrows}

\usepackage{color}
\usepackage{url}
\usepackage[colorlinks]{hyperref}
\hypersetup{%
        plainpages=true,
        breaklinks=true,
        hypertexnames=false,
        pageanchor=true,
        colorlinks=true,
        linkcolor={blue},
        citecolor={magenta},
        urlcolor={blue},
        anchorcolor={black}
      }
\usepackage[all]{hypcap} 
\usepackage[noabbrev]{cleveref} 

\usepackage{mleftright} 

\renewcommand{\eqref}[1]{\mbox{Eq.~(\ref{#1})}}

\newcommand{\figpanel}[2]{Fig.~\hyperref[#1]{\ref*{#1}(#2)}}
\newcommand{\figpanels}[3]{Fig.~\hyperref[#1]{\ref*{#1}(#2)-(#3)}}
\newcommand{\figpanelNoPrefix}[2]{\hyperref[#1]{\ref*{#1}(#2)}}

\usepackage[normalem]{ulem}

\begin{document}

\title{Forbidden transitions in superconducting artificial atoms}

\author{Alberto~Del~\'Angel}
\affiliation{Department of Microtechnology and Nanoscience,  Chalmers University of Technology, 41296 Gothenburg, Sweden}

\author{Th\'eo~S\'epulcre}
\affiliation{RIKEN Center for Quantum Computing, RIKEN, Wakoshi, Saitama 351-0198, Japan}

\author{Ricardo~Guti\'errez-J\'auregui}
\email{rg-jauregui@fisica.unam.mx}
\affiliation{Departamento de F\'isica Cu\'antica y Fot\'onica,  Instituto de F\'isica, Universidad Nacional Aut\'onoma de M\'exico, Ciudad de M\'exico, 04510, M\'exico}

\author{Anton~Frisk~Kockum}
\email{anton.frisk.kockum@chalmers.se}
\affiliation{Department of Microtechnology and Nanoscience,  Chalmers University of Technology, 41296 Gothenburg, Sweden}

\date{\today}

\begin{abstract}

Artificial atoms built from Josephson junctions have become a powerful tool to explore the limits of quantum optics due to their strong coupling to electromagnetic fields and their sensitivity to changes at the single-photon level. This sensitivity to quantum fluctuations complements their metrological and computational use, which are based on the precise oscillating frequency of the underlying supercurrents. We present here a theory for Josephson junctions immersed in electromagnetic fields where focus is shifted from temporal correlations and towards spatial ones. Unlike the commonly used circuit and black-box descriptions, our work is based on a microscopic model that enables systematically accounting for the effect of the spatial and vectorial profile of an electromagnetic field over a  junction. As an  example of the interactions that emerge in such a setup, we investigate the possibility of driving a junction via a quadrupole transition, using typical experimental parameters in existing devices. With the transition being dependent on the gradient of the electric field---rather than its intensity---the junction can be excited in a region where the electric field vanishes.

\end{abstract}

\maketitle

An atom probes its surrounding electromagnetic field by building correlations between its degrees of freedom and those of the field. This build-up follows strict conservation laws, which can be leveraged by tailoring the structure of the field for imaging~\cite{Chevy2000, Dorn2003, Rubinsztein-Dunlop2017}, control~\cite{Zwierlein2005, Hernandez-Rajkov2024, DelPace2022}, and explore the quantized exchange of dynamical variables~\cite{Andersen2006, Hernandez-Cedillo2013}. 

For artificial atoms built from superconducting circuits~\cite{Gu2017, Blais2020, Blais2021}, such as Josephson junctions, the coupling mechanism with electromagnetic fields is different. A Josephson junction is a macroscopic quantum system characterized by a phase and charge imbalance that summarize the highly correlated state of the underlying electrons~\cite{Martinis1987}. These degrees of freedom are still altered by external fields, which imprint their spatial and temporal structure to create a supercurrent that has been used to probe the quantized magnetic flux~\cite{Jaklevic1964} or acquire a metrological representation of the volt~\cite{Clarke1970}. The striking precision at which these devices work while being disordered at the microscopic level has been central for developing effective theories that describe their quantum behaviour and the construction of junction-based qubits for information processing and computation~\cite{Wendin2017, Kockum2019, Kim2023, Acharya2025}.

Superconducting artificial atoms are commonly described as lumped circuits~\cite{Bouchiat1998, Devoret2021}, ignoring the vectorial nature of the field and spatial profile of the junction, or using  self-consistent solutions of Maxwell equations~\cite{Nigg2012, Solgun_2014, Minev2021, Pham_2023, Lu2026}, blurring the line that separates the properties of field and atom. As fabrication and lithographic techniques continue to improve, recent experiments have reached a regime where the polarization and spatial profile of the field begins to be relevant. The precise location of the circuits inside waveguides and cavities or resonators has been exploited to explore decay paths~\cite{Houck2008} and create advanced measurement schemes~\cite{Hacohen-Gourgy2016, Sunada2022}. 

The dawn of this regime has also raised concerns about the modelling of the light-matter coupling, which gains importance as these systems scale in complexity and size. On the one hand, the coupling is non-trivial near boundaries where magnetic and electric fields strongly mix~\cite{Minev2021}. On the other hand, arrays of superconducting circuits can respond differently according to their relative location and the spatial profile of the electromagnetic environment mediating their interaction~\cite{Fink2009, Zanner2022}.

In this Letter, we present a systematic approach to model the interaction between a Josephson junction and an electromagnetic field that moves beyond the lumped-circuit and black-box descriptions. Our objective is to devise a theoretical scheme that allows us to explore how the vectorial and spatial profiles of the field are imprinted onto a junction. We begin from the Josephson relations, which describe the currents formed in a junction under an external field, written here in a way that emphasizes spatial effects. We then move towards a microscopic model where field and junction can be separated and quantum fluctuations be accounted for. This model allows us to present the transition rules that determine the sensitivity of the junction to a structured field. As a specific example, we show that, for realistic experimental parameters, a quadrupole transition can be leveraged to excite the junction in a location where the electric field is zero. Proper understanding of these effects provides a first step towards the use of structured light to control superconducting artificial atoms and, eventually, the creation of more complex structures constructed from several of these elements. Such advances can find applications in both quantum optics, quantum computing, and other quantum technologies.

\begin{figure}
\begin{center}
\includegraphics[width=.95\linewidth]{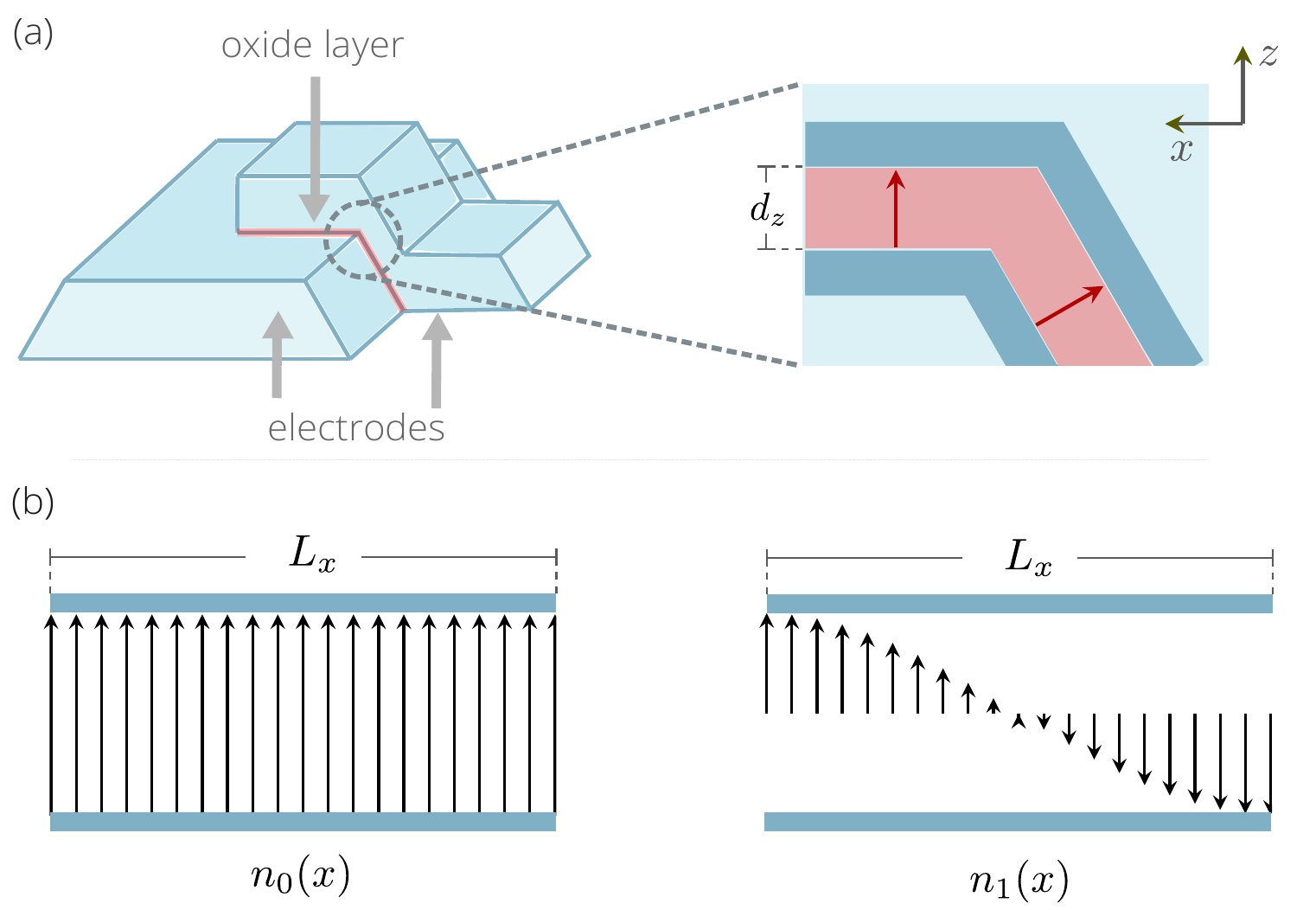}
\caption{Artificial atoms built from Josephson junctions couple to the surrounding electromagnetic environment. The coupling depends on the polarization and spatial profiles of field and junction.
(a) Sketch of a cross-strip junction~\cite{Devoret2021}. The oxide layer (light red; length $d_{z}\simeq \SI{1}{\nano\meter}$) allows for Cooper pairs to tunnel along two directions (red arrows) to create a polarization-sensitive coupling.
(b) Spatial profile of the charge imbalance for the fundamental and first mode of a flat junction ($L_{x} \simeq \qtyrange[range-phrase = -, range-units = single]{0.1}{10}{\micro\meter}$). For short junctions ($L_{x} \ll \lambda_{\text{J}}$), these modes can be weakly coupled by an external magnetic field to imbue the junction with a quadrupole moment.}\label{fig:proposal}
\end{center}
\end{figure}

\paragraph{Setup.}
The core of our study is a Josephson junction immersed inside an electromagnetic field. The junction is formed from two superconducting electrodes separated by a thin oxide layer, whose shape allows for electrons to tunnel along two main directions, as sketched in Fig.~\ref{fig:proposal}. The entire system is cooled below the critical temperature, where the electrons within each superconductor condense into a macroscopic state characterized by a phase $\theta_{\ell}(\mathbf{r})$ and Cooper-pair density $\rho_{\ell}(\mathbf{r})$ that spread along the volume $\mathcal{V}_{\ell}$ of the superconductor. We use the index $\ell = \lbrace 1,2 \rbrace$ to distinguish between the electrodes. 

It is the phase coherence of the condensate that determines the distinguishing traits of a superconductor and whose response to structured fields we are interested in. In the presence of an electromagnetic field described by scalar and vector potentials, $\phi$ and $\mathbf{A}$, respectively, the phase acts as a potential for the currents~\cite{Josephson1962, Werthamer1966, Anderson1967}:
\begin{equation}\label{eq:current_potential}
u_{\ell}(\mathbf{r}) = \nabla \theta_{\ell}(\mathbf{r}) -\frac{2e}{\hbar c} \mathbf{A}(\mathbf{r}) \, .
\end{equation}
Its evolution is described by the Josephson relations
\begin{subequations}\label{eq:dynamical_equation}
\begin{align}
\hbar \mathcal{V}_{\ell} \frac{ \partial \rho_{\ell}(\mathbf{r})}{\partial t} &= E_{\text{J}} \lambda_{\text{J}}^{2}\nabla \cdot u_{\ell}(\mathbf{r}) - E_{\text{J}} \sin \varphi_{\ell}(\mathbf{r})  \, , \\
\hbar \frac{ \partial \theta_{\ell}(\mathbf{r})}{\partial t} &= 2 E_{\text{C}} \mathcal{V}_{\ell} \rho_{\ell}(\mathbf{r}) - 2e \phi(\mathbf{r}) \, .
\end{align}
\end{subequations}
Here, $E_{\text{C}} = (2e)^2/2C$ is the Cooper-pair charging energy with $C$ the junction capacitance, while $E_{\text{J}}$ and $\lambda_{\text{J}}$ are the Josephson energy and length, respectively. The former represents the capacitive energy stored in the superconductors while the product of the latter two characterizes their kinetic inductance~\footnote{We have written the inductive energy stored in the superconductor in terms of the Josephson energy via the relation $\hbar^{2}\mathcal{V}/{8Le^{2}} = E_{J} \lambda_{J}$.}.

The Josephson relations allow us to distinguish effects in the bulk of the superconductors from those at the boundaries. From Eqs.~(\ref{eq:current_potential}) and (\ref{eq:dynamical_equation}), the Meissner effect and zero resistance that typify a superconductor follow from its successful attempt to smoothen spatial and temporal changes in the phase via screening currents. This smoothening means that, in order to control the response of the junction using fields of modest strength, we need to focus on the phase difference $\varphi$ that drives the supercurrent traversing the oxide layer. The phase difference depends on the vector potential along the shortest path connecting the electrodes,
\begin{equation}\label{eq:phase_diff}
\varphi_{1}(x,y) = \theta_{1}(x,y) - \theta_{2}(x,y) - \frac{2e}{\hbar c} \int_{z_1}^{z_2} \mathbf{A}(x,y,z)\cdot \mathbf{e}_{z} \text{d}z \, .
\end{equation}
When writing Eq.~(\ref{eq:phase_diff}) we have chosen this path along the $z$ axis to make explicit the difference between surface effects $\theta(x,y)$ from those inside the oxide layer. Moving forward, we denote the surface points as $\mathbf{r}_{s}$ with the electrons traversing the orthogonal path $\mathbf{l}$. Taking the inverse path, $\varphi=\varphi_{1}=-\varphi_{2}$.

By focusing on the phase difference and the accompanying charge imbalance $n(\mathbf{r}_{s}) = \tfrac12 \mathcal{V}[\rho_{1}(\mathbf{r}_{s})-\rho_{2}(\mathbf{r}_{s})]$, we can write the equations of motion in terms of the electric $\mathbf{E}$ and magnetic $\mathbf{B}$ fields inside the oxide layer. The Josephson relations then read
\begin{subequations}\label{eq:dynamical_equation_imbalance}
\begin{align}
\hbar \frac{ \partial n(\mathbf{r}_{s})}{\partial t} &= - E_{\text{J}} \sin \varphi(\mathbf{r}_{s})  + E_{\text{J}}\lambda_{\text{J}}^{2} \nabla \cdot \nabla \varphi(\mathbf{r}_{s})  \nonumber \\
&\quad + E_{\text{J}}\lambda_{\text{J}}^{2} \nabla \cdot \mleft[\frac{2e}{\hbar c}   \int_{}\text{d} \mathbf{l} \times \mathbf{B}(\mathbf{r}_{s},\mathbf{l}) \mright] \, , \\
\hbar \frac{ \partial \varphi(\mathbf{r}_{s})}{\partial t} &= 2 E_{\text{C}} n(\mathbf{r}_{s}) + 2e \int \text{d} \mathbf{l} \cdot \mathbf{E}(\mathbf{r}_{s},\mathbf{l}) \, .
\end{align}
\end{subequations}

\paragraph{Multipolar Hamiltonian.}
Equations~(\ref{eq:dynamical_equation}) and~(\ref{eq:dynamical_equation_imbalance}) accurately describe the behaviour of the junction---once complemented with the appropriate boundary conditions~\cite{Bulaevskii2006}. We, however, are interested in a theoretical scheme where the dynamics of the junction and electromagnetic field are differentiated. This scheme remains conducive to the equations of motion while allowing us to treat the case where quantum fluctuations become relevant using tools originally developed for neutral atoms inside cavities~\cite{Brune1996, CohenTannoudji1997, Carmichael2008}. To do so consistently, we present the details of a microscopic model in the Supplemental Material~\cite{SM_26}. The model begins from the weakly interacting electrons inside the electrodes and leads to a minimal-coupling Hamiltonian that describes the low-energy dynamics of the junction:
\begin{equation}\label{eq:minimal_coupling_Hamiltonian}
\hat{\mathcal{H}} = \hat{\mathcal{H}}_{1} + \hat{\mathcal{H}}_{2} - \frac{1}{\mathcal{V}_{\text{oxi}}}\int \text{d}V_{\text{oxi}} E_{\text{J}} \cos \hat{\varphi}   \, ,
\end{equation}
where the supercurrent lives inside the oxide layer of volume $\mathcal{V}_{\text{oxi}}$ and the free superconductors are described by
\begin{equation}
\hat{\mathcal{H}}_{\ell} = \int \frac{\text{d}V_{\ell}}{\mathcal{V}_\ell} E_{\text{C}}(\mathcal{V}_{\ell} \hat{\rho}_{\ell})^{2} + \frac{E_{\text{J}}\lambda_{\text{J}}^{2}}{2} \big(\nabla \hat{\theta}_{\ell} -\frac{2e}{\hbar c}\mathbf{A} \big)^{2}  - 2e \phi \mathcal{V}_{\ell} \hat{\rho}_{\ell} \, . 
\end{equation}
The phase and the charge density are conjugate variables that obey the commutation relation $[\exp(i \hat{\theta}_{\ell}(\mathbf{x})),\hat{\rho}_{\ell^\prime}(\mathbf{y})]=\exp(i \hat{\theta}_{\ell}(\mathbf{x})) \delta_{\ell,\ell^\prime} \delta(\mathbf{x}-\mathbf{y})$~\cite{Mukunda1979, Mandel1995, Devoret2007}. For coherent states the simplified relation $[\hat{\theta}_{\ell}(\mathbf{x}),\hat{\rho}_{\ell^\prime}(\mathbf{y})] = i \delta_{\ell,\ell^\prime} \delta(\mathbf{x}-\mathbf{y})$ is used~\cite{Anderson1967}. 

Using a gauge transformation~\cite{Mandel1979,CohenTannoudji1997}, Eq.~(\ref{eq:minimal_coupling_Hamiltonian}) turns into an effective, multipolar Hamiltonian:
\begin{align}\label{eq:multipole_coupling}
\hat{\mathcal{H}}_{\text{mp}} = \frac{1}{\mathcal{V}_{\text{oxi}}}&\int \text{d}V_{\text{oxi}} E_{\text{C}}\hat{n}^{2} - E_{\text{J}} \cos \hat{\varphi} + 2e \hat{n} \int \text{d}\mathbf{l} \cdot \mathbf{E}  \nonumber \\
& + \tfrac12 E_{\text{J}}\lambda_{\text{J}}^{2} \mleft( \nabla \hat{\varphi} + \frac{2e}{\hbar c} \int \text{d}\mathbf{l} \times \mathbf{B} \mright)^{2} \, .
\end{align}
that leads to the dynamical equations~(\ref{eq:dynamical_equation_imbalance}). The charge imbalance and phase difference written here represent the macroscopic degrees of freedom of the junction that satisfy equivalent commutation relations.

The multipolar Hamiltonian in Eq.~(\ref{eq:multipole_coupling}) is to be compared with the basic expressions derived from circuit theory~(see, for example, Refs.~\cite{Bouchiat1998, Devoret2007}). The former makes the spatial dependence of fields and junctions explicit, while the latter encompasses these traits inside effective quantities such as currents and voltages valid within the quasistatic regime underlying circuit theory~\cite{Collin1990, Blais2021}. Important differences, however, begin to appear when the junction is embedded in structured electromagnetic environments or placed near superconducting boundaries, where electric and magnetic fields change abruptly. In these regions, the microscopic model is better suited to deal with both the vectorial nature and spatial profile of the field. First, by allowing for tunneling along two orthogonal directions in the junction, we obtain a polarization-sensitive response. Such a response has been achieved in circuit theory by coupling two junctions~\cite{Minev2019, Minev2021}. Second, by maintaining the spatial overlap of fields and junction through the integrals of Eq.~(\ref{eq:multipole_coupling}), excitation paths depending on the shape of the field are made possible~\cite{Goetz_2018,Grankin_2024}. We now present a specific situation where this second, less explored distinction is important.

\paragraph{Imbuing a quadrupole moment.}
The selection rules that determine the excitations paths inside a junction provide an example of the spatial effects central to this work. For simplicity, we now focus on junctions with only one tunneling direction, where the main effects we wish to demonstrate are already present [see~\figpanel{fig:proposal}{b}]. Given the short distance separating the electrodes, $d_{z} \sim \SI{1}{\nano\meter}$,  the electromagnetic fields can be expanded around the center of the oxide layer $\mathbf{X}$. Up to second order, the resulting interaction Hamiltonian reads
\begin{equation}\label{eq:interaction}
\hat{\mathcal{H}}_{\text{int}} = {\mathbf{E}}(\mathbf{X})\cdot \hat{\mathbf{d}} + {\mathbf{B}}({\mathbf{X}})\cdot \hat{\mathbf{m}} + \tfrac12 \nabla{\mathbf{E}}({\mathbf{X}}):\hat{\mathbf{q}} \, ,
\end{equation}
where the electric dipole $\hat{\mathbf{d}}$, magnetic dipole $\hat{\mathbf{m}}$, and electric quadrupole $\hat{\mathbf{q}}$ operators represent the sensitivity of the junction to different properties of the field~\footnote{The couplings are: $\hat{\mathbf{d}} = 2e\int d \bar{V}_{\text{o}} \hat{n} \mathbf{r}_{\text{o}}$; $\hat{\mathbf{q}} = 2e\int d \bar{V}_{\text{o}} \hat{n} \mathbf{r}_{\text{o}}  \mathbf{r}_{\text{o}} $; $\hat{\mathbf{m}} = E_{\text{J}}\lambda_{\text{J}}^{2} (2e/\hbar c) \int d \bar{V}_{\text{o}} \mathbf{r}_{\text{o}}  \times \nabla \varphi$ with $d\bar{V}_{\text{o}} = \mathcal{V}_{\text{oxi}}^{-1} d{V}_{\text{oxi}}$.}. For the flat junction, the relevant components are
\begin{align}
&\hat{\mathbf{d}}_{z} = \mathcal{V}_{\text{oxi}}^{-1} (2ed_{{z}}) \int \text{d} {V}_{\text{oxi}} \hat{n}(\mathbf{r}_{s}) \mathbf{e}_{\mathbf{z}} \, , \label{eq:dipole_moment} \\
&\hat{\mathbf{q}}_{xz} = \mathcal{V}_{\text{oxi}}^{-1} (2ed_{z}) \int \text{d} {V}_{\text{oxi}} x \hat{n}(\mathbf{r}_{s}) \mathbf{e}_{x} \mathbf{e}_{z} \, , \label{eq:quadrupole_moment}\\ 
&\hat{\mathbf{m}}_{y} = \mathcal{V}_{\text{oxi}}^{-1}E_{\text{J}}\lambda_{\text{J}}^{2} (2e/\hbar c) d_{{z}} \int \text{d} {V}_{\text{oxi}} \mleft[ \mathbf{e}_{z} \times \nabla \hat{\varphi}(\mathbf{r}_{s}) \mright] \label{eq:mag_dipole_moment}\, .
\end{align}
Equations~(\ref{eq:dipole_moment})--(\ref{eq:mag_dipole_moment}) give unambiguous contributions from magnetic and electric field couplings to the junction.

The connection to the energy levels and selection rules of a radiating atom becomes apparent when the charge imbalance $\hat{n}$ is expanded within a normal-mode basis. A typical model to analyze these normal modes considers the junction as a transmission line driven by the supercurrent~\cite{Werthamer1966}. This picture leads to a feedback mechanism where the stationary states of line plus current follow Eq.~(\ref{eq:dynamical_equation_imbalance}) written as
\begin{equation}\label{eq:sine-gordon}
\nabla^{2} \varphi - (\lambda_{\text{J}} \omega_{0})^{-2} \partial_{t}^{2} \varphi = \lambda_{\text{J}}^{-2} \sin \varphi \, ,
\end{equation}
with the boundary conditions at the superconductors $\nabla {\varphi}\vert_{S} = (2e/\hbar c) (d_{z} + 2\lambda_{L}) \mathbf{e}_{\mathbf{l}} \times \mathbf{B}$~\cite{Anderson1967} and $ \hbar \omega_{0} = \sqrt{2 E_{\text{C}} E_{\text{J}}}$. The boundary condition reflects the currents caused by the magnetic field via the London penetration length $\lambda_{L}$. 

The fundamental mode of the junction has a constant phase, $\nabla^{2} \varphi = 0$. When excited, this mode displays a discrete non-linear spectrum whose energy levels are fixed by the Mathieu characteristic value~\cite{Cottet2002, Koch2007}. Its behaviour is simplified inside the transmon regime ($E_{\text{J}} \gg E_{\text{C}}$), where charge fluctuations become negligible. The supercurrent can be expanded around its zero-phase value in this regime and the energy levels resemble those of an anharmonic oscillator. Transitions between neighbouring levels are given by the lowering $\hat{b}_{0}$ and rising operators $\hat{b}_{0}^{\dagger}$ with~\cite{Koch2007}
\begin{subequations}\label{eq:quantum_oscillator-koch}
\begin{align}
\hat{n}_{0}(\mathbf{r}_{s}) & = (E_{\text{J}} /8E_{\text{C}} )^{1/4} (\hat{b}_{0} + \hat{b}_{0}^{\dagger}) \, , \\
\hat{\varphi}_{0}(\mathbf{r}_{s}) & = i (E_{\text{C}}/2 E_{\text{J}})^{1/4} (\hat{b}_{0} - \hat{b}_{0}^{\dagger}) \, .
\end{align}
\end{subequations}

Given the flat spatial profile of the fundamental mode, its quadrupole moment is zero [see Eqs.~(\ref{eq:quadrupole_moment}) and~(\ref{eq:quantum_oscillator-koch})] and quadrupole transitions through this mode are forbidden. Thus, in order to excite the junction via a quadrupole path, we need to consider other modes. In the transmon regime, these modes are
\begin{subequations}\label{eq:quantum_oscillator}
\begin{align}
\hat{n}_{m}(\mathbf{r}_{s}) & = \sqrt{\hbar \omega_{m}/4 E_{\text{C}}} \cos (k_{m}x) (\hat{b}_{m}+ \hat{b}_{m}^{\dagger}) \, , \\
\hat{\varphi}_{m}(\mathbf{r}_{s}) & = i  \sqrt{ E_{\text{C}}/\hbar\omega_{m}} \cos (k_{m}x) (\hat{b}_{m} - \hat{b}_{m}^{\dagger})\, .
\end{align}
\end{subequations}
They oscillate at the frequencies $\omega_{m} = \omega_{0} \sqrt{1 + \lambda_{\text{J}}^{2} k_{m}^{2}}$ with $\vert k_{m} \vert = \pi m/L_{x}$. Their parity follows that of $m$, such that the first excited mode drawn in \figpanel{fig:proposal}{b} has a vanishing dipole moment and a non-zero quadrupole one. 

While the higher modes can provide a quadrupole excitation path, they oscillate at prohibitely large frequencies for short junctions ($\lambda_{\text{J}} \gg L_{x}$). 
Taking an Nb-based junction with $\lambda_{\text{J}} = \SI{200}{\micro\meter}$, $L_{x} = \SI{10}{\micro\meter}$, and $\omega_0 / 2 \pi = \SI{7}{\giga\hertz}$~\cite{Martinis1987}, the first mode oscillates at $\omega_{1} / 2 \pi \simeq \SI{440}{\giga\hertz}$, below the superconducting gap. A junction, however, can still be imbued with a quadrupule moment near the fundamental frequency using a magnetic field orthogonal to the oxide layer ($\mathbf{B}=B \mathbf{e}_{y}$). This field modulates the phase difference across the junction~\cite{Eck1964, Coon1965, Barone1982}
\begin{equation}
\varphi' = \varphi + k_{B}x + \omega_{E} t
\end{equation}
with $k_{B} = 2e(2\lambda_{\text{L}} + d_{z})B/\hbar c$ and $\omega_{E} = 2ed_{} E  /\hbar $ to couple the bare modes. The coupling is readily obtained inside the transmon regime by expanding the supercurrent in Eq.~(\ref{eq:multipole_coupling}) around the modulated value and writing the quadratic phase within the normal-mode expansion of Eq.~(\ref{eq:quantum_oscillator}). The coupling strength between the fundamental and the $m$th mode is then
\begin{equation}
\hbar \Omega_{0,m} = \frac{\hbar \omega_{0}}{({1 + \lambda_{\text{J}}^{2} k_{m}^{2}})^{1/4}}\int_{0}^{L_{x}} \text{d}x \cos(k_{B}x) \cos(k_{m}x) \, .
\end{equation}
It is maximized when the magnetic modulation matches the target mode profile, a fact that has been exploited to measure the self-synchronization of the supercurrent in small junctions driven by oscillating fields~\cite{Coon1965}.

We use this self-synchronization to imbue a quadrupole moment. Using the resonance condition $k_{B} = k_{1}$~($B\simeq \SI{5}{}$G) only the fundamental and first-excited mode are strongly coupled. They give way to the dressed modes
\begin{subequations}\label{eq:normal_modes_quad}
\begin{align}
\hat{b}_{+} & = \cos \alpha \, \hat{b}_{1} + \sin \alpha \, \hat{b}_{0} \, , \\
\hat{b}_{-} & = \sin \alpha \, \hat{b}_{1} - \cos \alpha \, \hat{b}_{0} \, ,
\end{align}
\end{subequations}
with $\tan 2 \alpha = \Omega_{0,1} / (\omega_{0}-\omega_{1}) $~\footnote{The frequencies are normalized by the field (See SM).}. The lower mode oscillates near the fundamental frequency and displays a small quadrupole population $\sin \alpha \simeq 10^{-3} \cos \alpha$ for the parameters quoted above. While small, the excitation path opened by this quadrupole transition fundamentally changes the way to access the junction. Being sensitive to changes of the field---rather than its intensity---it becomes possible to excite the junction in regions where there is no electric field. We note that the idea to drive natural atoms inside dark regions was predicted in Ref.~\cite{Jauregui2004} using structured light and experimentally verified using a single trapped ion in Ref.~\cite{Schmiegelow2016}. 

The case of a junction coupled to a coplanar waveguide resonator~\cite{Wallraff2004, Fink2008} provides an illustrative example of the transition paths. The cosine profile of the second mode of this resonator displays large electric intensity at the antinodes and large gradients at the nodes, where the electric field vanishes. This competition between gradients and intensity is reflected in the ratio $R = 2\pi L_{x}/L_{\text{res}}$ between maximal quadrupole and dipole couplings. For a resonator of length $L_{\text{res}}= \SI{1}{\centi\meter}$ we obtain $R=2 \pi \times 10^{-3}$. Similar results follow for Al-based junctions~\cite{SM_26}. The main obstacle to observe the quadrupole transition is the product of this ratio and the quadrupole population of the dressed state. To overcome this obstacle a driven resonator or one with small photon-escape rate is needed. Both are within experimental reach~\cite{Blais2004, Haroche_2020}.  


\paragraph{Conclusions and outlook.}
We have presented a theoretical framework for the coupling between a Josephson junction and a structured electromagnetic field, placing emphasis on the spatial profile of these extended systems and their polarization effects. By accounting for these effects, we have showcased a situation where a junction is excited in a region with no electric field, a striking example of quadrupole transitions in superconducting artificial atoms. The theoretical framework, however, moves beyond this example and is constructed to describe a new generation of experiments with junctions immersed in complicated fields, as those found near the boundaries of superconducting surfaces, in a regime where quantum fluctuations of light and matter are relevant. 

The framework may find wide use in quantum optics and information processing with superconducting artificial atoms, where great precision in qubit design and high-fidelity performance is sought. By identifying situations where small deviations in the standard description begin to appear and quantifying its effects, we provide a path forward. Given the extraordinary success of circuit theory and its use for scaling, it would be interesting to present a systematic description of polarization and forbidden transitions in this theory. The quest for performance also meets fundamental debates about superradiant phase transitions~\cite{Nataf2010, Viehmann2011, Bamba2017, Stokes2022} and gauge~\cite{You2019, Kockum2019a, Stokes2022, Riwar_2022, Lu2026} in this platform, where our formalism can provide insights.

We have presented results for Nb- and Al-based junctions. These results should be extended to further accommodate experimental settings. Given the small dipole moment that follows the length of the oxide layer $d_{z}$, a superconducting artificial atom normally requires superconducting pads to couple to cavity fields. The effect of these pads over the field surrounding the junction is not considered in the calculations above.

\begin{acknowledgments}

\paragraph{Acknowledgments.}
We thank Vitaly Shumeiko and Roc\'io J\'{a}uregui for insightful discussions. RG-J acknowledges support from PAPIIT-UNAM (Grants No.~IA103024 and IA105926) and SECIHTI (Award No.~CBF-2025-I-1090). AAM and AFK acknowledge support from the Swedish Foundation for Strategic Research (Grant No.~FFL21-0279). AFK is also supported by the Swedish Foundation for Strategic Research (Grant No.~FUS21-0063), the Horizon Europe programme HORIZON-CL4-2022-QUANTUM-01-SGA via the project 101113946 OpenSuperQPlus100, and the Knut and Alice Wallenberg Foundation through the Wallenberg Centre for Quantum Technology (WACQT).

\end{acknowledgments}

\bibliography{References}

\end{document}


\title{Supplementary material for ``Forbidden transitions in superconducting artificial atoms''}

\author{Alberto~Del~\'Angel}
\affiliation{Department of Microtechnology and Nanoscience,  Chalmers University of Technology, 41296 Gothenburg, Sweden}

\author{Th\'eo~S\'epulcre}
\affiliation{RIKEN Center for Quantum Computing, RIKEN, Wakoshi, Saitama 351-0198, Japan}

\author{Ricardo~Guti\'errez-J\'auregui}
\email{rg-jauregui@fisica.unam.mx}
\affiliation{Departamento de F\'isica Cu\'antica y Fot\'onica,  Instituto de F\'isica, Universidad Nacional Aut\'onoma de M\'exico, Ciudad de M\'exico, 04510, M\'exico}

\author{Anton~Frisk~Kockum}
\email{anton.frisk.kockum@chalmers.se}
\affiliation{Department of Microtechnology and Nanoscience,  Chalmers University of Technology, 41296 Gothenburg, Sweden}

\date{\today}

\maketitle
\onecolumngrid

\tableofcontents
\vspace{1cm}

In this supplementary material, we run through a detailed description of the microscopic model that leads to the Hamiltonian description of the Josephson junction immersed in a structured electromagnetic field. We also present the experimental parameters required to realize our proposal using Nb- or Al-based Josephson Junctions.  For completeness and to connect with standard descriptions, this supplementary material is presented in SI units. The connection with the main text follows.

\section{Path-integral formulation}


\subsection{Minimal-coupling and tunneling Hamiltonians}

We consider a short Josephson junction formed by two aluminum superconducting electrodes separated by a thin oxide layer.
Our discussion begins from the minimal-coupling Hamiltonian describing the interaction between the electrons in a single superconductor $\hat{\psi}_{\sigma}(\vec{x})$ and both external and internal electromagnetic fields: 
\begin{equation}\label{Eq:MCHamiltonian}
    H_{ \text{MC} } = 
    \int \dtres x
    \mleft[ \;
    \sum_{\sigma} 
    \hat{\psi}^{\dagger}_{\sigma}
        \mleft\{    
        \frac{1}{2m}
            \mleft[
                \vec{p} - e ( \vec{A} + \vec{a} )
            \mright]^{2}
         + e \left[
                V + \phi
            \right]
        - \mu
        \mright\}
        \hat{\psi}_{\sigma}
        +
        \hat{\psi}^{\dagger}_{\uparrow} \hat{\psi}^{\dagger}_{\downarrow}
        \hat{\Delta} 
        + 
        \hat{\Delta}^{\dagger}
        \hat{\psi}_{\downarrow}
        \hat{\psi}_{\uparrow}
    \mright],
\end{equation} 
where $\sigma = \; \uparrow,\downarrow$ denotes the electron's spin up and down, respectively, $\mu$ is the chemical potential, and the integral extends over the superconductor's volume. 
In Eq.~(\ref{Eq:MCHamiltonian}), we have introduced the superconductor's electrostatic and magnetostatic potentials $V$ and $\vec{a}$, respectively, and the scalar and vector potentials $\phi$ and $\vec{A}$, respectively, that describe an external, and in general dynamic, electromagnetic environment.
The static potentials account for the capacitive and inductive energies stored in the superconductor and are obtained from the Coulomb and Biot-Savart interactions between electrons~\cite{jackson_classical_1999}:
\begin{subequations}
    \begin{align}
    W_{\text{E}} &=
    \frac{1}{8\pi\epsilon}
    \int \dtres x \dtres y \;
    \frac{ \rho(\vec{x}) \rho(\vec{y})
        }{
            \| \vec{x}-\vec{y} \|}
     \rightarrow \int \dtres x \;
     \frac{\epsilon}{2} \|\nabla V(\vec{x}) \|^{2}
     +
     \rho(\vec{x})V(\vec{x})
     ,\\
    W_{\text{M}} &=
     \frac{\mu_{0}}{8\pi} 
     \int\dtres x\dtres y \;
     \frac{\vec{J}(\vec{x})\cdot\vec{J}(\vec{y})
            }{ \| \vec{x}-\vec{y} \|}
     \rightarrow \int \dtres x \;
     \frac{1}{2\mu_{0}} \|\nabla \times \vec{a}(\vec{x}) \|^{2}
     +
     \vec{J}(\vec{x},\tau)\cdot\vec{a}(\vec{x}),
    \end{align}
\end{subequations}
where $\epsilon$  is the superconductor's permittivity, and $\rho = e \sum_{\sigma} \hat{\psi}^{\dagger}_{\sigma}\hat{\psi}_{ \sigma}$ and 
$
\vec{J} =
 e/2m \sum_{\sigma}
  [ 
    (\vec{p}\hat{\psi}^{\dagger}_{\sigma})
    \hat{\psi}_{\sigma}
     +
    \hat{\psi}^{\dagger}_{\sigma}
    (\vec{p} \hat{\psi}_{\sigma})
  ]
$ represent the charge and current densities, respectively.
The last two terms of Eq.~(\ref{Eq:MCHamiltonian}) describe the formation of the superconducting state through the pairing field $\hat{\Delta}(\vec{x})$.
Below the critical temperature, the system undergoes a spontaneous symmetry breaking and the pairing field acquires a non-trivial phase 
$
\hat{\Delta}(\vec{x})
 =
\Delta_{0}
    e^{
    i \hat{\theta}(\vec{x}) 
    }
$, $\Delta_{0} \in \mathbb{R}$, where $\hat{\theta}(\vec{x})$ encodes the macroscopic quantum coherence of the superconducting state.

In the presence of a sufficiently thin barrier, the electrons can tunnel between the two superconductors.
This process is described by a tunneling Hamiltonian
\begin{equation}\label{Eq:TunnelingHamiltonian}
    H_{\text{T}}
    = 
    \int \dtres x_{1} \dtres x_{2} \;
    \sum_{\sigma}
    t(\vec{x}_{1},\vec{x}_{2})
        \mleft[
            e^{
                i\chi_{ \text{P} }
                (\vec{x}_{1},\vec{x}_{2})
            }
            \hat{\psi}^{\dagger}_{\sigma 2}
            \hat{\psi}_{\sigma 1}
            + 
            \text{H.c.}
        \mright]
\end{equation}
where
$ 
            \chi_{\text{P}}(\vec{x}_{1},\vec{x}_{2}) =
            \frac{2e}{\hbar}
            \int_{ \vec{x}_{1} }^{ \vec{x}_{2}}
            \de \vec{l}
            \cdot 
            \vec{A}(\vec{l})
$
is the Peierls phase accounting for the coupling to the external electromagnetic field. 
We choose $\vec{l}$ to be the shortest path connecting points on each superconductor
$   \vec{l} 
    = 
    \vec{x}_{1}
    +
    u \left(
        \vec{x}_{2}
        -
        \vec{x}_{1}
    \right)
$
, $u \in [0,1]$.
To model the tunneling as local across the barrier, 
we propose the tunneling amplitude to be 
\begin{equation}
    t(\vec{x}_{1},\vec{x}_{2})
    = 
    t_{0} 
    \delta^{(2)}(\vec{x}_{s 1}-\vec{x}_{s2})
    \delta(z_{2} - z_{1} - d_{z})
    ,
    \;
    t_{0} \in \mathbb{R}^{+},
\end{equation}
where $\vec{x}_{s} = (x,y)$ are the transverse coordinates with respect to the tunneling axis $\hat{z}$.
This choice describes the tunneling across the barrier of thickness $d_{z}$ conserving the transverse coordinates during the tunneling process.

The appearance of the superconducting phase requires careful attention to the electromagnetic gauge invariance. 
Specifically, we can absorb the superconducting phase into the electron fields through a gauge transformation~\cite{RevModPhys.73.663}
\begin{subequations}\label{Eq:GaugeTransforms}
    \begin{align}
        \hat{\psi}_{\sigma}' & 
        = 
        e^{i
        \frac{\theta}{2}
        }
        \hat{\psi}_{\sigma}
        ,\\ 
        \vec{A}' & = \vec{A} + 
        \frac{\hbar}{2e}
        \nabla \theta, \\
        \phi' & =
        \phi 
        -
        \frac{\hbar}{2e}\partial_{t}\theta.
    \end{align}
\end{subequations}
Under these transformations, the superconducting phase is transferred to the kinetic, electric, and tunneling terms of the complete Hamiltonian:
\begin{align}
    H_{ \text{MC} } & = 
    \int \dtres x \;
    \hat{\Psi}^{\dagger}
    \mleft\{
    \mleft(
    \frac{1}{2m}
        \mleft[
            \vec{p}
            -
            e 
            \mleft(
             \frac{\hbar}{2e}\nabla \theta
             +
              \vec{A}
             +
             \vec{a}   
            \mright)
        \mright]^{2}
        - e 
        \mleft[
            \frac{\hbar}{2e}
            \partial_{t} \theta_{j}
            - 
            \phi
            -
            V
        \mright]
        -
        \mu
    \mright)\sigma_{z}
    +
    \Delta_{0}\sigma_{x}
    \mright\}
    \hat{\Psi}
    , \label{Eq:MCHamiltonian2} \\  
    H_{ \text{T} } & = 
    \int \de V_{\text{oxi}} \;
    t_{0} \mleft[
        \hat{\Psi}_{2}^{\dagger}
        e^{
            i\sigma_{z}
            \hat{\varphi}/2
           }
        \hat{\Psi}_{1}
        +
        \text{H.c.}
    \mright]
    , \label{Eq:TunnelingHamiltonian2}
\end{align}
where $\sigma_{z}$ and $\sigma_{x}$ are the Pauli $z$ and $x$ matrices, respectively,
$
\hat{\Psi}_{j} = 
    [
        \hat{\psi}_{\uparrow j}
        ,
        \hat{\psi}^{\dagger}_{\downarrow j}
    ]^{\intercal}
$
is the Nambu spinor describing the quasiparticles of the $j$th superconductor, and the phase in the tunneling Hamiltonian 
\begin{equation}
    \hat{\varphi}(\vec{x}) =
    \hat{\theta}_{2}(\vec{x}_{s}, z + d_{z})
    -
    \hat{\theta}_{1}(\vec{x}_{s}, z)
    -
    \frac{2e}{\hbar}
    \int 
    \de\vec{l}\cdot \vec{A}(\vec{l}),
\end{equation}
defines the gauge-invariant phase difference across the junction.


\subsection{Functional integral, gradient expansion, and coefficients of capacitance and inductance}

The low-energy dynamics of the system are governed by the superconducting phases introduced above.
To derive an effective theory of their coupling to the electromagnetic fields, it is convenient to employ a path-integral approach in which, since the quasiparticle fields in Eqs.~(\ref{Eq:MCHamiltonian2}) and (\ref{Eq:TunnelingHamiltonian2}) appear quadratically, they may be integrated out exactly.
This approach also allows us to systematically derive the coupling terms of the resulting theory and to make explicit the physical assumptions underlying the model.
Path-integral techniques have been widely used to derive effective descriptions of superconducting systems starting from microscopic Hamiltonians.
In particular, Refs.~\cite{Eckern1984Dec, Zazunov2003Feb} derive the Josephson effect and its generalization for weak-link junctions, ignoring the spatial variation of the phases and electromagnetic fields.
References~\cite{Altland2010Mar} and~\cite{Nagaosa1999} work out effective descriptions of the quantum electrodynamics of single superconductors in terms of their macroscopic degrees of freedom, retaining their spatial variations.
More recently, the path-integral approach has been used to study the quantum dynamics of superconducting qubits realized in a single, extended Josephson junction~\cite{Grankin2024}.

We now formulate the problem in terms of the imaginary-time partition function
\begin{equation}\label{Eq:PartitionFunction}
        \mathcal{Z}  =
        \int \mathcal{D}[\Psi] 
        \mathcal{D}
            [
            \theta,\vec{A},
            \phi,\vec{a},
            V
            ]
        \exp\mleft\{
            -\frac{1}{\hbar}
            \int_{0}^{\hbar \beta} \de\tau
            \mleft[
            \sum_{j=1}^{2}
            \mleft(
                \int \dtres x_{j} \;
                \bar{\Psi}_{j}
                \hbar \partial_{\tau}
                \Psi_{j}
                H_{\text{MC}j}
                +
                H_{\text{S}j}
            \mright)
                +
                H_{\text{T}}
            \mright]
            \mright\},
\end{equation} 
where $\beta = 1/k_{\text{B}}T$ and 
$
H_{\text{S}j }
= 
\int \dtres x_{j} 
\left[
\frac{\epsilon}{2} \| \nabla V_{j} \|^{2}
+
\frac{1}{2\mu_{0}} \| \nabla \times \vec{a}_{j} \|^{2} 
\right]
$
is the static electromagnetic energy stored in the $j$th superconductor.
Integrating out the fermionic fields yields an effective action 
$
S_{\text{eff}}
= - \hbar \text{Tr}\left[\log(G^{-1})\right] + S_{\text{S}}
$
, where $G$ is the Gor'kov Green's function including the electromagnetic potentials and tunneling terms, and $S_{\text{S}}$ contains the electrostatic contributions.
Expanding the effective action up to quadratic terms in the gradients of the phases and the electromagnetic potentials, and to second order in the tunneling amplitude, results in a low-energy theory
\begin{align}\label{Eq:EffectiveAction}
    S_{\text{eff}} &  = 
    \int \de \tau
    \mleft\{
        \sum_{j =1}^{2}
        \int  \dtres x_{j} \;
        i \frac{2e n_{0}}{\mathcal{V}}
        \mleft[
            V_{j} + \phi_{j}
        \mright]
        +
        \rho_{0}
        \mleft[
            \frac{\hbar}{2}
            \partial_{\tau}\theta_{j}
            +
            e( V_{j} + \phi_{j} )
        \mright]^{2}
        -
        \frac{ n_{0} }{2m\mathcal{V} }
        \mleft[
            \frac{\hbar}{2}\nabla\theta_{j}
            -
            e (\vec{a}_{j} 
            +
            \vec{A}_{j})
        \mright]^{2}
        \notag \mright.\\
        & \mleft.
        \quad \quad \quad \quad \; \;
         +
        \int 
        \frac{\de V_{ \text{oxi} } 
                }{
                    \mathcal{V}_{\text{oxi}}
                }
            E_{\text{J}}\cos\varphi
        + 
        \sum_{j =1}^{2}
        \int \dtres x_{j} \;
        \mleft[
            \frac{\epsilon}{2} \| \nabla V_{j} \|^{2}
            -
            \frac{1}{2\mu_{0}} \| \nabla \times \vec{a}_{j} \|^{2} 
        \mright] 
    \mright\}.
\end{align}
Here, $n_{0} = \sum_{k} [1- \xi_{k}/E_{k}]\tanh(\beta E_{k}/2)$ is the average number of Cooper pairs, 
$\mathcal{V}$ is the electrode volume, and $\rho_{0}$ denotes single-particle density of states evaluated at the Fermi level.
The Josephson energy is given by 
$
E_{\text{J}} = 
\mathcal{V}_{\text{oxi}}
\rho_{ \text{oxi} }
(t_{0}\Delta_{0})^{2}
\int_{-\infty}^{\infty} \de \xi 
\tanh(\beta E/2)/4E^{3}
$,
where
$
E = \sqrt{\xi^{2} + \Delta_{0}^{2} }
$
is the quasiparticle energy,
$
\xi^{2} = \hbar^{2}k^{2}/2m - \mu
$
is the free-particle energy with respect to the chemical potential, and 
$
\mathcal{V}_{\text{oxi}}
$
and
$
\rho_{\text{oxi}}
$
refer to the volume and density of states, respectively, in the junction region.
The vector potential is evaluated within each superconductor 
$
\vec{A}_{j} \equiv
\vec{A}(\vec{x}_{j},\tau)
$, and similarly for the scalar potential $\phi$.\\

The first term in Eq.~(\ref{Eq:EffectiveAction}) describes the coupling of the condensate charge density to the scalar potentials. 
In a neutral superconductor, this contribution is compensated by the ionic background charge density, resulting in a vanishing net monopole moment. 
Meanwhile, the static potentials $V_{j}$ and $\vec{a}_{j}$ account only for the electrostatic and magnetostatic response of the superconductors and do not represent independent dynamical degrees of freedom. 
We may consequently integrate them out of Eq.(~\ref{Eq:EffectiveAction}), yielding a redefinition of the coefficients appearing in the effective action:
\begin{equation}\label{Eq:EffectiveAction2}
    S_{\text{eff}} \rightarrow 
    \int \de \tau 
    \mleft\{
        \sum_{j = 1}^{2}
        \int \dtres x_{j}^{3} \;
        \mleft[
            \frac{C}{ \mathcal{V} (2e)^{2} }
            \mleft(
                \hbar \partial_{\tau} \theta_{j} + 2e \phi_{j}
            \mright)^{2}
            -
            \frac{1}{ 2 L' (2e)^{2} }
            \mleft(
               \hbar \nabla\theta_{j} - 2e \vec{A}_{j}
            \mright)^{2}
        \mright]
        + \int 
        \frac{  \de V_{\text{oxi} }
                }{
                    \mathcal{V}_{ \text{oxi} }
             }
        E_{\text{J}}\cos\varphi
    \mright\},
\end{equation} 
where $C$ and $L'$ are the effective coefficients of capacitance and kinetic inductance per unit length, respectively.

Below, we outline the procedure used to derive the effective capacitance.
We work in the Fourier domain using the convention 
$
f(\vec{x},\tau) = 1/\sqrt{ \hbar \beta \mathcal{V}}
\sum_{\boldsymbol{\lambda}}
f(\boldsymbol{\lambda}) e^{i[ \vec{x} \cdot \vec{Q} - \omega_{\lambda}\tau]}
$, where
$\boldsymbol{\lambda} = (\omega_{\lambda}, \vec{Q})
$.
We isolate the terms in Eq.~(\ref{Eq:EffectiveAction}) involving the electrostatic potential $V_{j}$.
Since $V_{j}$ appears quadratically in the effective action, we integrate it out to obtain a contribution in terms of the combination $\vartheta_{j} = \hbar\partial_{\tau}\theta_{j}/2 + e\phi_{j}$ only.
In the Fourier domain, this yields 
\begin{equation}
        S_{\text{E}}[\theta,\phi] = 
        \sum_{j \boldsymbol{\lambda}} 
        \frac{          
            \epsilon \rho_{0} e^{2} Q^{2}/2
            }{
            e^{2}\rho_{0} + \epsilon Q^{2}/2
            }
            \frac{      
                \vartheta_{j}(\boldsymbol{\lambda})
                \vartheta_{j}(-\boldsymbol{\lambda})
                }{
                e^{2}
                }.
\end{equation}
Transforming back to real space results in a spatially nonlocal kernel with a characteristic length given by the Thomas-Fermi screening length $\lambda_{\text{TF}} = \sqrt{\epsilon/2e^{2}\rho_{0} }$~\cite{Nagaosa1999}:
\begin{equation}
    S_{\text{E}}[\theta,\phi] =
    \sum_{j=1}^{2} 
    \int \dtres y_{j}\;
    \frac{\epsilon}{\lambda_{\text{TF}}} 
    \frac{          
            e^{ - \|\vec{y}_{j}\|/\lambda_{\text{TF} }    }
        }{
            8\pi \| \vec{y}_{j} \|
        }
    \int \de\tau \dtres x_{j}\;
    \frac{              
        \vartheta_{j}(\vec{x}_{j},\tau)
        \vartheta_{j}(\vec{x}_{j}+ \vec{y}_{j},\tau)
         }{
            \mathcal{V} (2e)^{2}
         }.
\end{equation}
In the regime where the phases vary slowly over this scale $\vert \nabla \theta_{j}\vert \ll \lambda_{\text{TF}}^{-1}$, we may regard the term $\vartheta_{j}$ approximately constant over the support of the kernel such that
$
\vartheta_{j}(\vec{x}_{j} + \vec{y}_{j}) 
\approx 
\vartheta_{j}(\vec{x}_{j})
$.
Thus, the action reduces to a local form
\begin{equation}
    S_{\text{E}} = 
    \sum_{j = 1}^{2} 
    \int \de\tau \dtres x_{j}\;
    \frac{ C_{j} }{(2e)^{2}} 
    \frac{  
        \mleft( 
            \hbar \partial_{\tau}\theta_{j} + 2e \phi_{j}
        \mright)^{2}
            }{
            \mathcal{V}
            },
\end{equation}
where $C_{j}$ defines the effective capacitance of the $j$th superconductor:
\begin{equation}
    C_{j} = 
    \int \dtres y_{j} \;
    \frac{\epsilon}{\lambda_{\text{TF}}} 
    \frac{          
            e^{ - \|\vec{y}_{j}\|/\lambda_{\text{TF} }    }
        }{
            8\pi \| \vec{y}_{j} \|
        }.
\end{equation}
In conventional electrostatics, capacitances are determined by the conductor's geometry through the Green's function of the Laplacian operator 
$\nabla_{\vec{x}}^{2}G(\vec{x},\vec{y})
 =
-\delta(\vec{x}-\vec{y})
$
supplied with the appropriate boundary conditions~\cite{Schwinger1998Sep}.
In our case, however, capacitances emerge from the screened Coulomb response of the condensate.
Since the two superconductors are identical, they are characterized by the same screening length and we may take their capacitances as equal, i.e., $C_{1} \approx C_{2} \equiv C$.

The magnetostatic contribution is obtained analogously by integrating out the internal vector potentials $\vec{a}_{j}$ in Eq.~(\ref{Eq:EffectiveAction}).
This results in an inductive term characterized by the inverse kinetic inductance per unit length
\begin{equation}\label{Eq:Inductance}
    \frac{1}{L'}  = 
    \frac{ e^{2} n_{0} }{m \mathcal{V} }
    \int \dtres y \;
    \frac{      
        e^{-\| \vec{y} \|/ \lambda_{ \text{L} } } 
         }{
           4\pi  \lambda_{ \text{L} }^{2} \| \vec{y} \|
         },
\end{equation} 
where
$
\lambda_{\text{L}} =
\sqrt{m\mathcal{V}/\mu_{0} e^{2} n_{0} }
$
is the London penetration length characterizing the screening of magnetic fields into the superconductors.


\section{Field Hamiltonian}

The effective action in Eq.~(\ref{Eq:EffectiveAction2}) describes the low-energy dynamics of the superconducting phases in each superconductor and across the junction, and is interpreted as a field theory in imaginary time. 
It has the standard form 
$
S = \int \de\tau \dtres x\; 
\mathcal{L}
$,
where the Lagrangian density depends on the superconducting phases and their derivatives 
$
\mathcal{L} \equiv 
\mathcal{L}(
        \theta_{j},
        \partial_{\tau}\theta_{j},
        \nabla\theta_{j}
           )
$.
This structure allows for a canonical formulation in terms of conjugate variables.
We define the conjugate momentum to the superconducting phase in real time $t$ as
\begin{equation}
\rho_{j}(\vec{x},t) = 
\frac{  \partial \mathcal{L}
        }{
        \partial \mleft[
                \hbar \partial_{t}\theta_{j}
                 \mright]
     }
     = 
     \frac{2C}{(2e)^{2}}
     \frac{ 
            \hbar \partial_{t}\theta_{j} + 2e \phi_{j}
          }{ 
         \mathcal{V}
          }.    
\end{equation}

We note that the action in Eq.~(\ref{Eq:EffectiveAction2}) depends on gauge-invariant combinations of the superconducting phase and the electromagnetic potentials.
Under a gauge transformation
$
\theta_{j}' = \theta_{j} + \Lambda_{j}
$,
$
\phi_{j}' = \phi_{j} - (\hbar/2e) \partial_{t}\Lambda
$,
the combination $\hbar\partial_{t}\theta_{j} + 2e\phi_{j}$ remains invariant. 
As a consequence, the conjugate variable defined above is itself gauge-invariant and therefore corresponds to a physically observable quantity, which we identify as the charge density of the superconducting state in the $j$th superconductor. 
In terms of the pairs of conjugate variables, we obtain the Hamiltonian of the system via a Legendre transform
$
 \mathcal{H} = 
    \sum_{j}
    \int \dtres x_{j} 
    \mleft[
        \rho_{j}
         \mleft(
             \hbar \partial_{t}\theta_{j}
         \mright)
    \mright]
    -
    L
$:
\begin{equation}\label{Eq:HamiltonianCoulomb}
    \hat{\mathcal{H}} = 
    \sum_{j = 1}^{2}
    \int \frac{\dtres x_{j}}{\mathcal{V}}
    \mleft[
        E_{ \text{C} } 
        \mleft(
            \mathcal{V}\rho_{j}
        \mright)^{2}
        -
        2e \phi_{j}
        \mleft(
            \mathcal{V} \rho_{j}
        \mright)
        +
        \frac{ \hbar^{2} \mathcal{V} }{ 2 (2e)^{2} L' }
        \mleft(
        \nabla \theta_{j} - \frac{2e}{\hbar} \vec{A}_{j}
        \mright)^{2}
    \mright]
    -
    \int \frac{\de V_{\text{oxi} }}{\mathcal{V}_{\text{oxi}}} \;
    E_{\text{J}}
    \cos \varphi,
\end{equation}
where $E_{\text{C}} = (2e)^{2}/2C$ is the charging energy of a single Cooper pair.
The first three terms in Eq.~(\ref{Eq:HamiltonianCoulomb}) are characteristic of the minimal-coupling Hamiltonian of charged particles in an electromagnetic field written in Coulomb's gauge \cite{Cohen-Tannoudji:113864}, where the Coulomb energy
$
\int \dtres x_{j} E_{\text{C}} \mathcal{V}\rho_{j}^{2}
$
now acts as the kinetic energy term of the $j$th superconductor, while the 
$
\mleft(
    \vec{p}-2e\vec{A}
\mright)^{2}/2m
$ 
term is replaced by the superconducting phase gradient 
$
\|
\nabla\theta_{j} - (2e/\hbar) \vec{A}_{j}
\|^{2}
$.

Applying canonical quantization to the theory, we promote the fields to operators satisfying the equal-time commutation relations
\begin{equation}
    \mleft[
        \hat{\theta}_{j}(\vec{x})
        ,
        \hat{\rho}_{k}(\vec{y})
    \mright]
    = 
    i\delta_{jk}\delta(\vec{x}-\vec{y}).
\end{equation} 
This relation should be understood in the context of the effective theory we derived, where we treat the superconducting phase as a continuous field and the charge fluctuations are large compared to the single-electron scale.
At the microscopic level, however, the compact nature of the superconducting phase implies a discrete charge spectrum, and the above commutation relation should be modified to 
$
\left[
    \exp[i
    \hat{\theta}_{j}(\vec{x})]
    ,
    \hat{\rho}_{k}(\vec{y})
\right]
= \exp[i\hat{\theta}_{j}(\vec{x})]\delta_{jk}\delta(\vec{x}-\vec{y})
$~\cite{RevModPhys.40.411}.


\section{Equations of motion and gauge transformations: Multipolar Hamiltonian}

Using the equations of motion for the individual phases and densities computed from the Hamiltonian in Eq.(\ref{Eq:HamiltonianCoulomb}), we can construct the equations obeyed by the junction's degrees of freedom.
These are the `bare' phase difference 
$
\hat{\Theta}(\vec{x}) = 
\hat{\theta}_{2}(\vec{r}_{s},z+d)
-
\hat{\theta}_{1}(\vec{r}_{s},z)
$
and the charge imbalance operator 
$
\hat{n}(\vec{x}) = 
(\mathcal{V}/2)
    \left[
        \hat{\rho}_{2}(\vec{r}_{s},z+d)
        -
        \hat{\rho}_{1}(\vec{r}_{s},z)
    \right]
$:
\begin{subequations}\label{Eq:EOMSCoulomb}
    \begin{align}
            \frac{\partial \hat{\Theta}(\vec{x})}{\partial t} & =
            2 E_{\text{C}} \hat{n}(\vec{x})
            -
            2e \int \de \vec{l} \cdot \nabla \phi_{j}(\vec{l})
            ,\\
            \frac{\partial \hat{n}(\vec{x})}{\partial t} & = 
            - E_{\text{J}} \sin \hat{\varphi}(\vec{x}) 
            + 
            \frac{ \mathcal{V} }{2(2e)^{2}L'}
            \nabla \cdot 
                \mleft[
                    \nabla \hat{\Theta}(\vec{x})
                    -
                    2e \int \mleft(
                         \de \vec{l}\cdot \nabla 
                            \mright)
                            \vec{A}(\vec{l})
                \mright].
    \end{align}
\end{subequations}
These equations are general in the sense that they are invariant under gauge transformations and that there is a clear separation between matter $(\hat{\Theta} , \hat{n})$ and field $(\phi,\vec{A})$ degrees of freedom.
We may obtain a clearer physical picture of these equations by appropriately adding gradients of the Peierls phase 
$
\chi_{\text{P}} = 
(2e/\hbar) \int \de\vec{l}\cdot\vec{A}(\vec{l})
$:
\begin{subequations}\label{Eq:EOMSPoincare}
    \begin{align}
        \hbar \frac{\partial \hat{\varphi}(\vec{r}_{s})}{\partial t}  & = 
        2 E_{\text{C}} \hat{n}(\vec{r}_{s}) 
        +
        2e \int \de \vec{l} \cdot \vec{E}(\vec{l},\vec{r}_{s})
        ,\\
        \hbar \frac{\partial \hat{n}(\vec{r}_{s})}{\partial t} & = 
        - E_{\text{J}} \sin\hat{\varphi}(\vec{r}_{s}) 
        +
        E_{\text{J}}\lambda_{\text{J}}^{2}
        \nabla \cdot 
            \mleft[
                \nabla \hat{\varphi}(\vec{r}_{s})
                + 
                \frac{2e}{\hbar}\int \de \vec{l} \times \vec{B}(\vec{l},\vec{r}_{s})
            \mright],
    \end{align}
\end{subequations}
where $
\lambda_{ \text{J} } = 
\mleft[
    \phi_{0}/2\pi \mu_{0} (d+ 2\lambda_{\text{L}})
    J_{\text{c}}
\mright]^{-1/2}
$
is the Josephson penetration length obtained from the kinetic inductance Eq.~(\ref{Eq:Inductance}) and the Josephson energy jointly.
The explicit appearance of the electric and magnetic fields in Eqs.~(\ref{Eq:EOMSPoincare}) implies that including the Peierls phase in the definition of the phase difference $\Theta \rightarrow \varphi$ is equivalent to performing a gauge transformation into the multipolar gauge~\cite{10.1119/1.1491265}:
\begin{subequations}
    \begin{align}
        \hat{\varphi} & =
        \hat{\Theta} + \chi_{\text{P}}
        ,\\
        \phi'  & =
        -
        \int \de \vec{l} \cdot \vec{E}(\vec{l},\vec{r}_{s})
        ,\\
        \vec{A}' & =
        -
        \int \de \vec{l} \times \vec{B}(\vec{l},\vec{r}_{s}).
    \end{align}
\end{subequations}
This result parallels the Power--Zienau--Woolley transformation
$
\chi_{\text{PZW}} = 
e\int_{0}^{1} \de \lambda \;
\vec{x} \cdot \vec{A}( \lambda \vec{x})
$ 
typically employed for describing the quantum light-matter interactions of molecular and atomic systems in terms of their multipole moments coupled explicitly to the electromagnetic fields rather than the scalar and vector potentials~\cite{PhysRevResearch.2.013206}.

We may alternatively derive Eqs.~(\ref{Eq:EOMSPoincare}) from the multipolar Hamiltonian
\begin{equation}\label{Eq:HamiltonianMP}
    \hat{\mathcal{H}}_{\text{MP}} =
    \int \frac{\de V_{\text{oxi} } }{\mathcal{V}_{\text{oxi} } }
    \mleft[
        E_{\text{C}} \hat{n}^{2}
        -
        E_{ \text{J} } \cos\hat{\varphi}
        +
        2e \hat{n} \int \de\vec{l}\cdot\vec{E}(\vec{l})
        +
        \frac{  
            E_{\text{J}}\lambda_{\text{J}}^{2}
            }{
            2
            }
        \mleft(
            \nabla \hat{\varphi}
            +
            \frac{ 2e }{ \hbar }
            \int \de\vec{l}\times\vec{B}(\vec{l})
        \mright)^{2}
    \mright].
\end{equation}
In addition to the dynamical equations for the phase difference and the charge imbalance, the gauge-invariant phase difference satisfies a geometric constraint relating its spatial variation at the junction interface to the magnetic field.
This follows from its definition in terms of a line integral over the vector potential and reads~\cite{Barone1982Jul}
\begin{equation}\label{Eq:GradMagField}
    \nabla \hat{\varphi}\vert_{S} = 
    \frac{2e}{\hbar}(d + 2\lambda_{ \text{L} })
    \mleft( \hat{z} \times \vec{B}\mright)\vert_{S}.
\end{equation}


\section{Normal modes of a Josephson junction}

We want to treat the interaction between the junction and the external electromagnetic field as a perturbation.
To do this, we obtain a normal-mode decomposition of the junction degrees of freedom by diagonalizing the quadratic part of the junction Hamiltonian in absence of the electromagnetic fields:
\begin{equation}\label{Eq:JunctionHamiltonian}
    \hat{\mathcal{H}}_{0} =
    \int \frac{\de V_{ \text{oxi} }}{\mathcal{V}_{\text{oxi} }}
    \mleft[
             E_{\text{C}} \hat{n}^{2}
             +
             \frac{E_{\text{J}} \lambda_{\text{J}}^{2} }{2}
             \mleft(
                \nabla \hat{\varphi}
             \mright)^{2}
             -
             E_{\text{J} }
             \cos \hat{\varphi}
    \mright].
\end{equation} 
We decouple the equations of motion for the phase difference and the charge imbalance computed from the above Hamiltonian in the regime of small oscillations to obtain the linearized sine-Gordon equation 
\begin{equation}\label{Eq:LinearSineGordon}
    \lambda_{\text{J}}^{2}\nabla^{2}\hat{\varphi}
    - 
    \frac{1}{\omega_{0}^{2}}
    \partial_{t}^{2} \hat{\varphi}
    - \hat{\varphi}
    = 0, 
    \quad 
    \nabla\hat{\varphi}\vert_{S} = 0,
\end{equation}
where $\omega_{0} = \sqrt{2E_{\text{C}}E_{\text{J}}}/\hbar$ is the Josephson plasma frequency.
We consider a planar junction with sides $L_{x}$, $L_{y}$ perpendicular to the tunneling axis.
The solutions of Eq.~(\ref{Eq:LinearSineGordon}) considering vanishing derivatives at the junction's interface are known as the Swihart modes~\cite{Swihart1961Mar}
\begin{subequations}\label{Eq:NormalModes}
    \begin{align}
        \hat{\varphi}(x,y) & =
        i \sum_{\vec{k}} 
        \sqrt{
                \frac{ E_{\text{C} }
                    }{
                    \hbar \omega_{ \vec{k} } 
                    }
             }
             f_{\vec{k} }(x,y)
             \mleft[
                \hat{b}_{\vec{k} } - \hat{b}_{ \vec{k} }^{\dagger}
             \mright]
             ,\\
        \hat{n}(x,y) & = 
        \sum_{\vec{k}}
        \sqrt{      \frac{      
                        \hbar\omega_{\vec{k} }
                    }{
                        4 E_{\text{C}}
                    }
             }
            f_{\vec{k}}(x,y)
            \mleft[
                \hat{b}_{\vec{k} } + \hat{b}_{ \vec{k} }^{\dagger}
            \mright]
            ,\\
        f_{\vec{k}}(x,y) & = 
        A_{x}A_{y}
        \cos(k_{x}x)\cos(k_{y}y).
    \end{align}
\end{subequations}
Here, the wave vector $\vec{k} = \pi [n/L_{x},m/L_{y}], \; n,m \in \mathbb{N}$ is related to the mode's natural frequency through the dispersion relation 
$
\omega_{\vec{k}} = \omega_{0}
\sqrt{ 1 + \vert \vec{k} \vert^{2}\lambda_{\text{J}}^{2}     }
$
and the $\hat{b}_{\vec{k}}, \hat{b}_{\vec{k}}^{\dagger}$ are bosonic operators satisfying the commutation relation
$
\mleft[
\hat{b}_{\vec{k}},\hat{b}_{\vec{q} }^{\dagger}
\mright] 
= 
\delta_{\vec{k}\vec{q}}
$.
The spatial profiles of each mode are normalized with respect to the inner product
$
\int_{0}^{L_{x}}(\de x/L_{x})
\int_{0}^{L_{y}}(\de y/L_{y})
f_{\vec{k}}f_{\vec{q}}
= \delta_{ \vec{k} \vec{q} }
$,
so that 
$A_{x,y} \equiv A_{n,m} = \sqrt{2}, \; n,m > 0$ and $A_{0} = 1$.



\subsection{Effect of a constant magnetic field}

We consider the effect of an applied constant static magnetic field $B_{0}$ along the $\hat{y}$ axis on the junction degrees of freedom.
From Eq.~(\ref{Eq:GradMagField}), the boundary condition for the gradient along the $\hat{x}$ axis becomes 
$
\partial_{x}\hat{\varphi}\vert_{0,L_{x}} 
=
k_{B}
$,
with $k_{B} = (2e/\hbar)(d+2\lambda_{\text{L} })B_{0}$.
We handle the boundary condition by splitting the solution of Eq.~(\ref{Eq:LinearSineGordon}) $\hat{\varphi}' = \phi_{0} + \hat{\varphi}$ into a classical background solution that absorbs the boundary condition $\phi_{0} = k_{B}x$ and a fluctuating part $\hat{\varphi}$ satisfying
$
\nabla\hat{\varphi}\vert_{S}  = 0
$.
Plugging this solution back into the full sine-Gordon equation and expanding the sine term to first order in $\hat{\varphi}$ yields an equation of the form 
\begin{equation*}
    \lambda_{\text{J}}^{2} \nabla^{2} \hat{\varphi}
    -
    \frac{1}{\omega_{0}^{2}}\partial_{t}^{2} \hat{\varphi}        
    -
    \cos(\phi_{0})\hat{\varphi}
    =
   \sin\phi_{0}.
\end{equation*}

Thus, the magnetic field turns the linearized sine-Gordon equation into a driven inhomogeneous equation for which no simple solutions exist in general, except when the oscillating terms are truncated to first order in $\phi_{0}$.
Yet, we can see the influence of the magnetic field on the system using the normal modes we previously computed by looking at the Josephson potential 
\begin{equation}
        \hat{\mathcal{H}}_{\text{J}}  = 
        - E_{\text{J}}
        \int \frac{ \de V_{ \text{oxi} }    }{\mathcal{V}_{\text{oxi}}}
        \cos\mleft(
            \phi_{0}
            +
            \hat{\varphi}
            \mright)
        = 
        - E_{\text{J}}
        \int \frac{ \de V_{ \text{oxi} }    }{\mathcal{V}_{\text{oxi}}}
        \mleft(
            \cos\phi_{0} \cos\hat{\varphi}
            -
            \sin\phi_{0} \sin\hat{\varphi}
        \mright).
\end{equation}
As we will see below, the first term in the above Hamiltonian is responsible for the mixing of the junction modes, whereas the second acts as a non-linear driving source. 
Expanding to second order in $\hat{\varphi}$, the quadratic part of the first term reads
\begin{equation*}
    \hat{\mathcal{H}}_{\text{J}}^{(2)} = 
    \sum_{\vec{k}\vec{q}}
    \frac{  \hbar \omega_{0}^{2} }{
            \sqrt{
                 \omega_{\vec{k}}\omega_{\vec{q}}
                 }
         }
         I_{\vec{k}\vec{q}}
    \mleft[
        \hat{b}_{\vec{k}}^{\dagger}\hat{b}_{\vec{q}}
        +
        \hat{b}_{\vec{k}} \hat{b}_{\vec{q}}^{\dagger}
        -
        \hat{b}_{\vec{k}}^{\dagger} \hat{b}_{\vec{q}}^{\dagger}
        -
        \hat{b}_{\vec{k}} \hat{b}_{\vec{q}}
    \mright],
\end{equation*}
where
$
I_{\vec{k}\vec{q}} =
\delta_{k_{y} q_{y}} 
A_{k_{y}}^{2} 
\int_{0}^{L_{x}} (\de x/L_{x})
A_{k_{x}} A_{q_{x}}
\cos(k_{B} x)
\cos(k_{x}x)
\cos(q_{x}x)
$
is the overlap integral whose oscillatory structure is characteristic of interference phenomena. 

Ignoring the contributions that do not conserve the number of excitations, the quadratic part of the junction Hamiltonian in Eq.~(\ref{Eq:JunctionHamiltonian}) becomes
\begin{equation}\label{Eq:MagneticHamiltonian}
    \hat{\mathcal{H}}_{0}^{(2)} =
    \sum_{\vec{k}}
    \frac{\hbar \omega_{\vec{k} } }{2}
    \mleft(
        1+ I_{\vec{k}\vec{k}}
    \mright)
    \mleft[
        \hat{b}_{\vec{k}}^{\dagger} \hat{b}_{\vec{k}}
        +
        \frac{1}{2}
    \mright]
    +
    \sum_{\vec{k} \neq \vec{q} }
    \frac{  \hbar \omega_{0}^{2} 
         }{
            \sqrt{
                \omega_{\vec{k}}\omega_{\vec{q}}
            }
         }
         I_{\vec{k}\vec{q}}
    \mleft[
         \hat{b}_{\vec{k}}^{\dagger} \hat{b}_{\vec{q}}
         +
         \hat{b}_{\vec{k}} \hat{b}_{\vec{q}}^{\dagger}
    \mright].
\end{equation}
The diagonal overlaps renormalize the Swihart modes' resonant frequency, while the off-diagonal overlaps induce coherent couplings between distinct modes. From the structure of the latter, we obtain that the applied magnetic field couples different modes that vary along the $\hat{x}$ axis only. 
If we consider the coupling of the fundamental and first excited modes $\hat{b}_{0}$, $\hat{b}_{1}$,  $\vec{k}_{1} = \pi[ 1/L_{x},0] $, Eq.~(\ref{Eq:MagneticHamiltonian}) becomes a $2\times2$ matrix
\begin{subequations}
    \begin{align}
    \frac{ \hat{ \mathcal{H} }_{0}^{(2)} }{\hbar} & = 
    \begin{bmatrix}
        \hat{b}_{0}^{\dagger} & \hat{b}_{1}^{\dagger}
    \end{bmatrix}
    \begin{bmatrix}
        \omega_{0}' & \Omega \\
        \Omega & \omega_{1}'
    \end{bmatrix}
    \begin{bmatrix}
        \hat{b}_{0}\\
        \hat{b}_{1}
    \end{bmatrix}
    \notag ,\\
       \omega_{0}' & = 
       \frac{ \omega_{0} }{2}
       \mleft[
            1
            +
            \text{sinc}(k_{B}L_{x})
       \mright]
      \notag ,\\
       \omega_{1}' & = 
       \frac{\omega_{1}}{2}
        \mleft(
           1 + \text{sinc}(k_{B}L_{x})  
           +
           \frac{1}{2}
           \mleft[
            \text{sinc}([ k_{B} - k_{1} ]L_{x})
            +
            \text{sinc}([ k_{B} + k_{1} ]L_{x})
           \mright]  
        \mright)
        \notag ,\\
        \Omega & = 
        \frac{  \omega_{0}
             }{
              \sqrt[4]{
                1 + k_{1}^{2}\lambda_{ \text{J} }^{2} 
                }
             }
        \frac{1}{\sqrt{2}}
        \mleft[
            \text{sinc}([k_{B} - k_{1}]L_{x})
            +
            \text{sinc}([k_{B} + k_{1}]L_{x})
        \mright]
        , \label{Eq:ModeCoupling}
    \end{align}
\end{subequations}
with $\text{sinc}(x) = \sin(x)/x$.\\

Diagonalizing the reduced Hamiltonian yields a pair of dressed operators $\hat{b}_{\pm}$ with natural frequencies 
\begin{equation}
\omega_{\pm} =
(1/2)(\omega_{1}'-\omega_{0}')
\pm
\sqrt{
    (\omega_{1}'-\omega_{0}')^{2}/4
    +
    \Omega^{2}
}.
\end{equation}
These relate to the bare modes $\hat{b}_{0,1}$ via a unitary matrix 
\begin{equation}
M = 
\begin{bmatrix}
    \cos\alpha & -\sin\alpha\\
    \sin\alpha & \cos\alpha
\end{bmatrix}
,
\end{equation}
with 
$
\cos\alpha = 
\frac{\Omega}
{
    \sqrt{
    \omega_{-}^{2} + \Omega^{2}
    }
}
$
and
$
\sin\alpha = 
\frac{ \omega_{-}}
{
    \sqrt{
    \omega_{-}^{2} + \Omega^{2}
    }
}
$
.
Due to the small coupling between the bare modes, the resonant frequency of the lower mode can be approximated as
$
\omega_{-} \approx 
\omega_{0}'
[    1
     +
     (L_{x}/\lambda_{\text{J} })^{2}
]
$
.
For the experimentally reported values $L_{x}/\lambda_{\text{J}} = 0.02$, the frequency difference between the fundamental and lower modes is on the order of \SI{0.1}{\mega\hertz}. 
For the upper mode we obtain
$
\omega_{+} \approx
\omega_{1}' + \omega_{0}'(L_{x}/\lambda_{\text{J}})^{2}
$.
Since $L_{x} \ll \lambda_{\text{J}}$, the mixing between the bare modes of the junction remains weak and the dressed natural frequencies remain close to their bare counterparts. 
Therefore, the excitation of the lower dressed mode involves only a small contribution of the first excited mode. 
The gauge-invariant phase difference and charge imbalance operators in terms of the dressed operators read
{\small
\begin{subequations}
    \begin{align}
        \hat{\varphi}(x,y) & = 
        i \hat{b}_{-}
        \mleft[
            \sqrt{
                \frac{E_{\text{C} }}{\hbar\omega_{0} }
                }
                \cos\alpha
                -
            \sqrt{2}\sqrt{
                \frac{E_{\text{C}}}{\hbar\omega_{1} }
            }\cos(k_{1}x)\sin\alpha
        \mright]
        +
        i\hat{b}_{+}
         \mleft[
            \sqrt{
                \frac{E_{\text{C} }}{\hbar\omega_{0} }
                }
                \sin\alpha
                +
            \sqrt{2}\sqrt{
                \frac{E_{\text{C}}}{\hbar\omega_{1} }
            }\cos(k_{1}x)\cos\alpha
        \mright]
        +
        \text{H.c.}
        ,\\
        \hat{n}(x,y) & = 
         \hat{b}_{-}
        \mleft[
            \sqrt{
                \frac{\hbar\omega_{0} }{ 4E_{\text{C} } }
                }
            \cos\alpha
            -
            \sqrt{2}\sqrt{
               \frac{\hbar\omega_{1} }{ 4E_{\text{C} } }
            }
            \cos(k_{1}x)
            \sin\alpha
        \mright]
        +
        \hat{b}_{+}
         \mleft[
            \sqrt{
                \frac{\hbar\omega_{0} }{ 4E_{\text{C} } }
                }
                \sin\alpha
                +
            \sqrt{2}\sqrt{
                \frac{\hbar\omega_{1} }{ 4E_{\text{C} } }
            }\cos(k_{1}x)\cos\alpha
        \mright]
        +
        \text{H.c.}
    \end{align}
\end{subequations}
}


\section{Multipole moments of a small Josephson junction}

We now characterize the electrodynamics of the lower dressed mode in terms of its multipole moments.
To do so, we project the multipolar Hamiltonian Eq.~(\ref{Eq:HamiltonianMP}) onto $\hat{b}_{-}$ and perform a multipole expansion of the electromagnetic fields around a reference point $\vec{X}$ inside the junction. 
The electric part of Eq.~(\ref{Eq:HamiltonianMP}) is 
\begin{equation}\label{Eq:ElectricIntHamiltonian}
    \hat{\mathcal{H}}_{\text{E}} =
    2ed \int \frac{\de V_{\text{oxi}}}{\mathcal{V}_{\text{oxi}}}
    \hat{n}(x,y)
    \int_{0}^{1} \de u \;
     E_{z}(\vec{l},x,y).
\end{equation}
Expanding the electric field around $\vec{X}$ and computing the integral with respect to the straight-line parameter $u$ generates the electric multipole couplings.
The leading contribution corresponds to the electric dipole interaction
\begin{equation*}
    \hat{\mathcal{H}}_{\text{E1}} = 
    2ed \int \frac{\de V_{\text{oxi}}}{\mathcal{V}_{\text{oxi}}} \;
    \hat{n}(x,y)
    E_{z}(\vec{X}).
\end{equation*}
Performing the volume integral over the spatial mode of the lower mode yields its electric dipole moment $\mathbf{d}_{z}$:
\begin{subequations}
    \begin{align}
          \hat{\mathcal{H}}_{\text{E1}} & = 
          \mleft[
            \hat{b}_{-}
            +
            \hat{b}^{\dagger}_{-}
          \mright]
          \mathbf{d}_{z}E_{z}(\vec{X})
          ,\\
          \mathbf{d}_{z} & =
          2ed
          \sqrt[4]{
            \frac{E_{\text{J} }}{8 E_{\text{C} }}   
          }\cos\alpha. 
    \end{align}
\end{subequations}
Thus, the lower dressed mode inherits the same electric dipole moment of the fundamental Swihart mode $\hat{b}_{0}$ up to corrections given by the mixing angle $\cos\alpha$.

The next contribution corresponds to the electric quadrupole interaction
\begin{equation*}
    \hat{\mathcal{H}}_{\text{E2}}  = 
    2ed
    \int \frac{\de V_{\text{oxi}}}{\mathcal{V}_{\text{oxi}}} \;
    \hat{n}(x,y)
    \left(r_{sj} - X_{j}\right)
    \partial^{j}E_{z}(\vec{X}),
\end{equation*}
where we have used the Einstein sum convention over repeated indices.
Projecting onto the lower dressed mode yields
\begin{subequations}
    \begin{align}
        \hat{\mathcal{H}}_{\text{E2}} & = 
        \mleft[
            \hat{b}_{-}
            +
            \hat{b}^{\dagger}_{-}
        \mright]
        \mathbf{q}_{zx}\partial_{x}E_{z}(\vec{X})
        ,\\
        \mathbf{q}_{zx} & = 
        \frac{\sqrt{2}}{\pi^{2}}
        2e d L_{x}
        \sqrt{
                \frac{\hbar\omega_{1} }{ 4E_{\text{C} } }
             } 
        \sin\alpha .
    \end{align}
\end{subequations}
Unlike the dipole moment, the quadrupole moment is proportional to $\sin\alpha$, and therefore vanishes in the absence of the external magnetic field.
It is thus clear that its origin comes from the mixture between the fundamental and first excited bare modes into the lower dressed mode.

Finally, we consider the magnetic term of the multipolar Hamiltonian
\begin{equation*}
    \hat{\mathcal{H}}_{\text{M1}} =
    \frac{ E_{\text{J} } \lambda_{\text{J}}^{2} }{\hbar} 
    \int \frac{\de V_{\text{oxi}}}{\mathcal{V}_{\text{oxi}}} \;
    2e d
    \nabla \hat{\varphi}
    \cdot
    \int_{0}^{1} \de u\;
    \mleft[
        \hat{z}
        \times
        \vec{B}(\vec{l})
    \mright],
\end{equation*}
and obtain the magnetic dipole interaction when remaining to first order in the expansion of the magnetic field:
\begin{subequations}
    \begin{align}
        \hat{\mathcal{H}}_{\text{M1}} & = 
        i\mleft[
            \hat{b}_{-}
            -
            \hat{b}^{\dagger}_{-}
        \mright]
        \mathbf{m}_{y}B_{y}(\vec{X}),\\
        \mathbf{m}_{y} & = 
        \frac{ E_{\text{J} } \lambda_{\text{J}}^{2} }{\hbar} 
        \frac{2ed}{L_{x}}
        \sqrt{
                \frac{2 \hbar\omega_{1} }{ E_{\text{C} } }
             }
        \sin\alpha.
    \end{align}
\end{subequations}
As in the electric quadrupole case, the magnetic dipole moment arises entirely from the hybridization of the bare modes of the junction. 
We thus conclude that the presence of an external magnetic field enriches the electromagnetic response of the junction to electromagnetic fields even if the mode mixing is weak.

\section{Experimental considerations}
The relevant parameter that determines the strength of the coupling $\Omega$ between the fundamental and first excited Swihart modes is controlled by the ratio $\lambda_{\text{J}}/L_{x}$ [see Eq.~(\ref{Eq:ModeCoupling})].
While increasing the lateral dimension $L_{x}$ enhances the coupling, excessively large junctions are undesirable in quantum electrical circuits since quantum coherence is progressively lost in the macroscopic limit~\cite{PhysRevB.35.4682}. 
Thus, it is preferable to engineer the Josephson penetration length through the system's material and fabrication parameters
\begin{equation}
    \lambda_{\text{J}} = 
    \sqrt{\frac{    
            \phi_{0}
            }{
            2\pi \mu_{0} 
            (d_{z} 
            +
            2\lambda_{\text{L}} )
            J_{c}
            }}.
\end{equation}
This length depends both on the electromagnetic properties of the superconducting electrodes, through the London penetration length $\lambda_{\text{L}}$, and on fabrication-dependent junction parameters through the critical current density $J_{\text{c}}$.
These quantities provide two experimentally accessible knobs for maximizing the hybridization between modes.
We now discuss representative values of the Josephson length, the natural frequency of the first excited mode $\omega_{1}$, and the strength of the magnetic field $B_{0}$ needed to target its spatial mode profile $k_{1} = k_{B}$.\\

For thin-film Al Josephson junctions, the effective London penetration length ranges from \SI{60}{\nano\meter} to \SI{160}{\nano\meter}, where the highest value is reached for the lowest film thickness of \SI{30}{\nano\meter}~\cite{Lopez-Nunez2025Sep}.
For Nb-based junctions, effective penetration lengths on the order of \SI{100}{\nano\meter} are reported for thin films of comparable thickness~\cite{PhysRevB.57.14416} (in the main text we consider the larger penetration length \SI{200}{\nano\meter} obtained in the seminal work of Ref.~\cite{Martinis1987} to display the robustness of our procedure). Meanwhile, critical current densities in both Al- and Nb-based junctions  are commonly found in the range of \qtyrange[range-phrase = -, range-units = single]{0.1}{1}{\milli\ampere/\micro\meter\squared}~\cite{Tolpygo2017Feb}.
Using representative parameters for Al: $J_{c} = \SI{0.5}{\milli\ampere/\micro\meter\squared}$
and $\lambda_{\text{L}} = \SI{100}{\nano\meter}$, we obtain a Josephson penetration length $\lambda_{\text{J}} \approx \SI{50}{\micro\meter}$.
For a junction with lateral side
$L_{x} = \SI{5}{\micro\meter}$
we obtain $\lambda_{\text{J}}/L_{x} = 10$, placing the system in a regime of appreciable but perturbative mode hybridization.
In this regime, and for $\omega_{0}/2\pi = \SI{6}{\giga\hertz}$, the first excited Swihart mode remains well below the Cooper-pair breaking energy $2\Delta_{0}$ for both Al and Nb junctions ($\hbar \omega_{1}/2\Delta_{0\text{Al}} = 0.75$, $\hbar \omega_{1}/2\Delta_{0\text{Nb}} =0.09$) and can in principle be addressed without quasiparticle generation.
Finally, the magnetic field required to match the wavenumber of the first excited mode is on the order of $B_{0} = \phi_{0}/4L_{x}\lambda_{\text{L}} = 10\:\mathrm{G} = \SI{e-3}{\tesla}$, which lies well within experimentally accessible field strengths and does not adversely affect the superconductivity for thin-film Al or Nb.

\bibliography{References_SM.bib}